# Importance and Techniques of Information Hiding : A Review

Richa Gupta[1], Sunny Gupta[2], Anuradha Singhal[3]

[13](Department of Computer Science, University of Delhi, India)
[2](University of Delhi, India)

***ABSTRACT*** **:** *Information or data is very crucial resource to us. Thus securing the information becomes all the more necessary. The communication media through which we send data does not provide data security, so other methods of securing data are required. Information hiding plays a very crucial role today. It provided methods for encrypting the information so that it becomes unreadable for any unintended user. This paper reviews the techniques that exist for data hiding and how can these be combined to provide another level of security.*

## 1. INTRODUCTION

Data or information is very crucial to any organization or any individual person. None of us likes our conversation being overheard as it contains the potential of being misused. Same is the case with the data of any organization or of any person. The exchange of data among two potential parties must be in done in a secured method so as to avoid any tampering. Two types of threats exists during any information exchange. The unintended user who may try to overhear this conversation can either tamper with this information to change its original meaning or it can try to listen to the message with intention to decode it and use it to his/her advantage. Both these attacks violated the confidentiality and integrity of the message passed. Providing intended access and avoiding unintended access is a very challenging task. Information hiding has been since long time. In past, people used hidden pictures or invisible ink to convey secret information [1].

## 2. IMPORTANCE

The importance of data hiding techniques comes from the fact the there is no reliability over the medium through which the information is send, in other words the medium is not secured. So, some methods are needed so that it becomes difficult for unintended user to extract the information from the message. Few reasons behind data hiding are:

1. Personal and private data
2. Sensitive data
3. Confidential data and trade secrets
4. To avoid misuse of data
5. Unintentional damage to data, human error and accidental deletion of data
6. Monetary and blackmail purposes
7. To hide traces of crime
8. Lastly, for fun

## 3. FEATURES FOR INFORMATION HIDING TECHNIQUES

Any information hiding technique shall exhibit certain characteristics:

1. Capacity – capacity refers to the amount of information that can be hidden in cover medium [1]. The amount of information that can be hidden is governed by the fact that information hidden should not completely alter the original message, in order to avoid the attention of unintended user.

2. Security – the information hiding method should provide security for data such that only the intended user can gain access to it. In order words, it refers to the inability of un-authorized user to detect hidden information. This is very crucial to protect the confidentiality and sensitivity of information being sent [1, 2].
3. Robustness – it refers to the amount of information that can be hidden without showing any negative effects and destroying hidden information [1].
4. Perceptibility – the data hiding method should hide data in such a manner that the original cover signal and the hidden data signal are perceptually indistinguishable.

## 4. DATA HIDING TECHNIQUES

There are three major data hiding techniques popular: watermarking, cryptography and steganography

### 4.1 WATERMARKING

A watermark is a recognizable image or pattern that is impressed onto paper, which provides evidence of its authenticity [3, 4]. Watermark appears as various shades of lightness/darkness when viewed in transmitted light. Watermarks are often seen as security features to banknotes, passports, postage stamps and other security papers [4]. Digital watermarking is an extension of this concept in the digital world.

Today there have been so much of data over internet that it has forced us to use mechanisms that can protect ownership of digital media. Piracy of digital information is very common, be it images, text, audio or video. These can be produced and distributed very easily. So, it becomes very important to find out who is the owner of the document. Digital watermarking provides a solution for longstanding problems faced with copyright of digital data [3]. Digital watermark is a kind of marker covertly embedded to any digital data such as audio or image data. It can later be extracted or detected to make assertion about data. This information can be information about author, copyright or an image itself [3, 5]. The digital watermark remains intact under transmission/transformation, allowing us to protect our ownership rights in digital form. Digital watermarks are only perceptible under certain conditions, i.e. after using some algorithm, and imperceptible anytime else. If a digital watermark distorts the carrier signal in a way that it becomes perceivable, it is of no use.

A watermarking system's primary goal is to ensure robustness, i.e, it should be impossible to remove the watermark without tampering the original data. Digital watermarking is a passive protection tool. It just marks the data, but does not degrade it nor controls access to data [5].

One application of digital watermarking is *source tracking*. A watermark is embedded into a digital signal at each point of distribution. If a copy of the work is found later, then the watermark may be retrieved from the copy and the source of the distribution is known. This technique reportedly has been used to detect the source of illegally copied movies. Another application is in broadcast monitoring, the television news often contains watermarked video from international agencies.

### 4.2 CRYPTOGRAPHY

Crypt means "hidden or secret" and graphein means "writing". The term has been derived from Greek language. Cryptography is an art of transforming data into an unreadable format called cipher text. The receiver at other side, deciphers or decrypt the message into plain text. Cryptography provides data confidentiality, data integrity, authentication and non-repudiation. Confidentiality is limiting access or placing restriction on certain types of information. Integrity is maintaining and assuring the accuracy of data being delivered, i.e, information contains no modification, deletion etc. Authentication ensures the identity of sender and receiver of the information. Non-repudiation is the ability to ensure that the sender or receiver cannot deny the authenticity of their signature on the sending information that they originated [6].

Modern age cryptography is synonymous with encryption. Here the original information is known as plain text and encrypted information is known as cipher text. Cryptography has 3 steps:

1. Encryption – encrypting the plain text to some non-readable form. The output of this is known as cipher text.
2. Message transfer – it involves sending the cipher text to the recipient.
3. Decryption – the receiver at the other side, decrypts the cipher text to get the original plain text.

Cryptography can be broadly categorized into symmetric key cryptography and asymmetric key cryptography.

1. Symmetric key cryptography – it refers to the encryption methods in which the sender and the receiver share the same key. Many encryption algorithms like AES, DES, RC5 etc uses this method of encryption.

Symmetric key cryptography has five components: plain text, encryption algorithm, secret key, cipher text and decryption algorithm. Encryption algorithm performs various operations on plain text using secret key. Secret key is independent of plain text and is chosen by one of the communicating parties. The output produced by this is cipher text. Decryption algorithm takes cipher text and secret key as input and produce plaintext as output.

A significant disadvantage of symmetric key cipher is that it requires the secret key to shared by each pair of communicating parties, and also the key itself to be shared in a secured medium. Any unintended user having the secret key possesses a threat of ciphering the text.

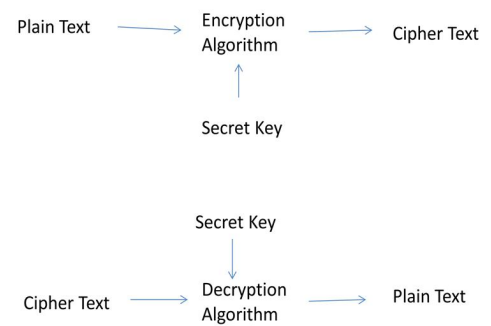

**Figure 1: Symmetric Key Cryptography**

2. Asymmetric key cryptography - it is also known as public key cryptography. It uses two keys: public key and private key. Public key can be freely distributed while its pairing private key must be kept secret. The public key is used or encryption. The cipher text is then sent to receiver. At the receiver end, it uses secret key with decryption algorithm to get the plain text.

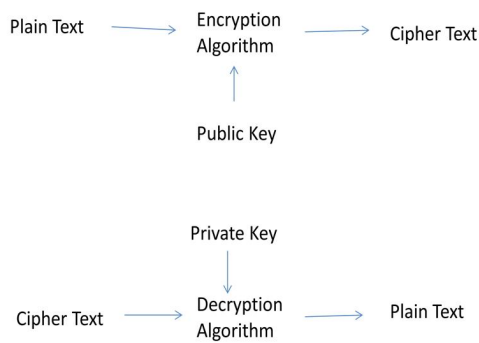

**Figure 2: Assymetric Key Cryptography**

Public key cryptography can also be used in digital signatures. Digital signatures can be permanently tied to the content of the message being signed. Secret key is used for signing the contents and the corresponding public key is used to validate the authenticity of the signature.

### 4.3 STEGANOGRAPHY

Steganography is a practice of hiding/concealing the message, file, image within other message, file or image. The word steganography is of Greek origin and means "covered writing" or "concealed writing" [7]. In other words, it is the art and science of communicating in a way which hides the existence of the communication. The goal is to hide messages inside other harmless messages in a way that does not allow enemy to even detect that there is a second message present [3]. Steganography focuses more on high security and capacity. Even small changes to stego medium can change its meaning. Steganography masks the sensitive data in any cover media like images, audio, video over the internet. Steganography involves four steps:

1. Selection of the cover media in which the data will be hidden.
2. The secret message or information that is to be masked in the cover media.
3. A function that will be used to hide data in the cover media and its inverse to retrieve the hidden data.
4. An optional key or the password to authenticate or to hide and unhide the data [2].

The cover chosen should be done very carefully. The cover chosen should contain sufficient redundant information which can be used to hide the data, because steganography works by replacing the redundant data with the secret message.

There are three basic types of steganographic protocols.

1. Pure steganography – it does not require the exchange of cipher such as a stego-key but the sender and receiver must have access to embedding and extraction algorithm. The cover for this method is c hosen such that it minimizes the changes caused by embedding process. These systems are not very secure as the security depends on the presumption that no other party is aware of this secret message.
2. Secret key steganography – this method uses a key to embed the secret message into the cover. The key is only known to sender and the receiver and is known prior to communication. Also, the key should be exchanged in a secure medium. The disadvantage of this approach is that it is susceptible to interception.

3. Public key steganography – it uses two keys, public key stored in public database and is used for embedding process and the secret key is known only to communication parties and is used to reconstruct the original message [8].

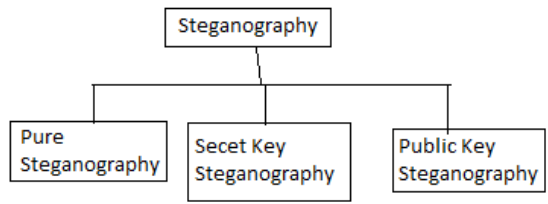

Figure 3: Steganographic Protocols

Types of steganographic techniques:
1. Text steganography
2. Audio and video steganography
3. Image steganography
4. IP steganography

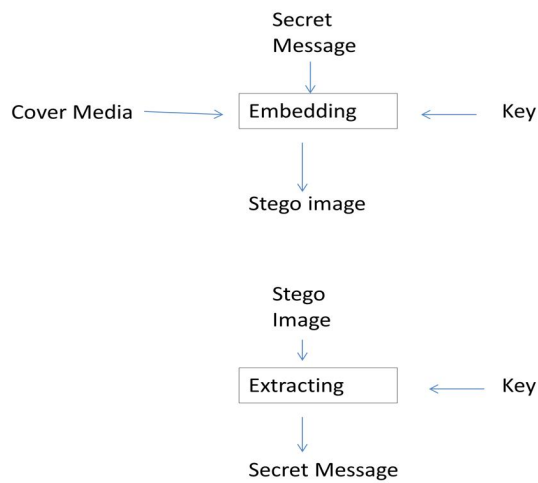

Figure 4: Steganography

## 5. STEGANOGRAPHY VS. CRYPTOGRAPHY

Steganography means "cover writing" whereas cryptography means "secret writing" [3]. Steganography is often confused with cryptography but there is substantial difference amongst two. The former uses a cover to hide the information and send it to the network. It is difficult for any unintended user to determine whether there is any secret information embedded or not. The important characteristic with steganography is that the cover should be chosen with enough redundant information so that even after embedding the message, it is not easy to detect for the message after looking at the message. Whereas, cryptography involves encrypting the message such that either it becomes unreadable or the original meaning of the message is entirely changed.

Steganography does not alter the structure of the secret message whereas, cryptography alters the structure of the secret message. Former prevents the discovery of the existence of the communication whereas latter prevents unauthorized user from discovering the contents of a communication

## 6. COMBINED CRYPTOGRAPHY AND STEGANOGRAPHY

Both the techniques can be combined to provide one more level of protection. The message can be first encrypted using cryptography to a cipher text. This cipher text then can be embedded In a cover media using steganography. This combined approach will satisfy the three goals of data hiding: security, capacity, robustness.

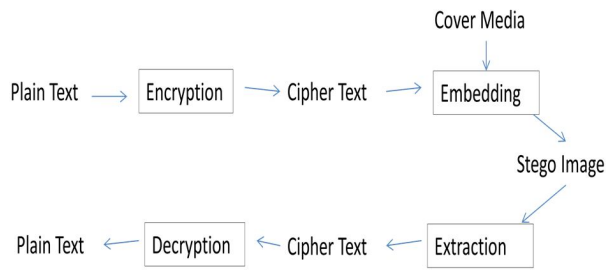

Figure 5: Cryptography with Steganography

## 7. COMBINED WATERMARKING AND STEGANOGRAPHY

To protect the authenticity of the document, watermarking can be applied to it. This watermarked document can be embedded in cover image using a stego-key and transmitted over the communication medium. At the receiver end, the information can be first decrypted using the reverse procedure and then it can be validated for its authenticity using the watermarking. This combined approach will satisfy all four goals of data hiding: security, capacity, robustness and perceptibility.

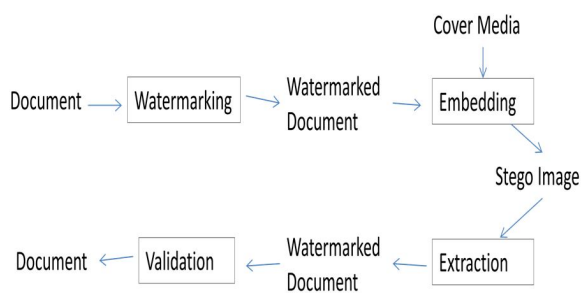

Figure 6: Watermarking with Steganography

## 8. CONCLUSION

In this paper, we have tried to give a review of existing data hiding techniques, their advantages and disadvantages. This paper also tells why data hiding is gaining importance these days and the goals that must be achieved of any data hiding technique. Also, we have tried to state how the basic goals of data hiding can be achieved by combining one or more techniques of data hiding.